\documentclass[12pt]{article}
\usepackage{setspace}
\usepackage{epsfig}
\usepackage{amssymb}
\usepackage{graphicx}
\def\be{\begin{equation}}
\def\ee{\end{equation}}
\def\ba{\begin{array}}
\def\ea{\end{array}}
\def\beqn{\begin{eqnarray}}
\def\eeqn{\end{eqnarray}}
\def\nonum{\nonumber}
\def\bt{\begin{tabular}}
\def\et{\end{tabular}}
\def\bc{\begin{center}}
\def\ec{\end{center}}

\begin{document}

\title{Leptonic mixing angle $\theta_{13}$ and ruling out of minimal texture for Dirac neutrinos}

\author{Priyanka Fakay, Samandeep Sharma, Gulsheen Ahuja$^*$, Manmohan
Gupta\\ {\it Department of Physics, Centre of Advanced Study,
P.U.,
 Chandigarh, India.}\\
\\{\it $^*$gulsheenahuja@yahoo.co.in}}

 \maketitle

\begin{abstract}

Implications of recently measured leptonic mixing angle
$\theta_{13}$ as well as the other two mixing angles have been
examined for Fritzsch-like mass matrices with minimal texture for
Dirac neutrinos. Interestingly, the existing data seems to rule
out this texture specific case of Dirac neutrinos for normal,
inverted hierarchy as well as degenerate scenario of masses.
\end{abstract}

\section{Introduction}
The recent measurements \cite{t2k}-\cite{reno} regarding the
neutrino mixing angle $\theta_{13}$ have undoubtedly improved our
knowledge of neutrino oscillation phenomenology. Interestingly,
this $\theta_{13}$ value which is unexpectedly `large', being
almost near the Cabibbo angle, would have important implications
for flavor physics. Also, it may be mentioned that before the
measurement of $\theta_{13}$, assuming it to be zero or nearly
equal to zero and considering the canonical values of the other
two neutrino mixing angles, the effort was to discover some
underlying symmetry \cite{sym} in the leptonic sector. The non
zero value of $\theta_{13}$ leads to parallelism between the
mixings of quarks and leptons as well as signifies the difference
between the mixing angles of quarks and leptons as the leptonic
mixing angles are large compared with the corresponding quark
mixing angles.

Ever since the observations regarding $\theta_{13}$ there has been
a good deal of activity on the theoretical front in understanding
the pattern of neutrino masses and mixings. Noting that there is a
similarity between quark and lepton mixing phenomena
\cite{smirnov}, it becomes desirable to understand these from the
same perspective as far as possible. However, there are some
important differences which have to be kept in mind before
considering a unified framework for formulating quark and lepton
mass matrices on the same footing. For example, one may note that
unlike the case of quark mixings which show a hierarchical
structure, the pattern of neutrino mixings do not show any
explicit hierarchy. Further, at present there is no consensus
about neutrino masses which may show normal/inverted hierarchy or
may even be degenerate. Furthermore, the situation becomes
complicated when one realizes that it is yet not clear whether
neutrinos are Dirac or Majorana particles.

It may be mentioned that in the absence of any viable theory for
flavor physics, one usually resorts to phenomenological models. In
this context, texture specific mass matrices have got good deal of
attention in the literature, for details and extensive references
we refer the readers to a recent review article \cite{ourreview}.
In particular, Fritzsch-like texture specific mass matrices seem
to be very helpful in understanding the pattern of quark mixings
and CP violation \cite{firstfri, quagroup}. Keeping in mind quark
lepton parallelism \cite{smirnov} and taking clue from the success
of these texture specific mass matrices in the context of quarks,
several attempts \cite{firstfri, neutgroup} have also been made to
consider similar lepton mass matrices. However, noting the above
mentioned complexities of neutrino masses and mixings, it seems
necessary to carry out a detailed and case by case analysis of
texture specific mass matrices for their compatibility with the
mixing data. In particular, for any given texture, the analysis
needs to be carried out for all the neutrino mass hierarchies as
well as for both Majorana and Dirac neutrinos since the latter
have not yet been ruled out experimentally \cite{diracstrumia}.

Considering neutrinos to be Majorana particles, after the recent
measurements of $\theta_{13}$, a few analyses have been carried
out for texture specific mass matrices in the non flavor basis. In
particular, Fukugita {\it et al.} \cite{fukugita} have
investigated the implications of angle $\theta_{13}$ on minimal
texture mass matrices (Fritzsch-like texture 6 zero) for normal
hierarchy of masses. This analysis has been extended further by
Fakay {\it et al.} \cite{our56majnew} wherein for all the
hierarchies of neutrino masses, texture 6 and 5 zero mass matrices
have been examined in detail. For the case of Dirac neutrinos,
although several authors have examined the possibility of these
having small masses \cite{smalldirmasses} as well as their
compatibility with the supersymmetric GUTs \cite{christoph},
however, similar attempts have not yet been carried out after the
measurements of $\theta_{13}$. In this context, it may be added
that the original texture 6 zero Fritzsch mass matrices have been
ruled out in the case of quarks, therefore, in the light of
similarity between the mixing patterns of quarks and leptons, it
becomes desirable to examine similar mass matrices for the cases
of neutrinos.

In the present paper, for the case of Dirac neutrinos, we have
carried out detailed calculations pertaining to mass matrices with
minimal texture for the three possibilities of neutrino masses
having normal/inverted hierarchy or being degenerate. In
particular, the analysis has been carried out by imposing
Fritzsch-like texture 6 zero structure on Dirac neutrino mass
matrices as well as on charged lepton mass matrices. The
compatibility of these texture specific mass matrices have been
examined by plotting the parameter space corresponding to the
recently measured mixing angle $s_{13}$ along with the other two
mixing angles $s_{12}$ and $s_{23}$. Further, for the normal
hierarchy case, the implications of mixing angles on the lightest
neutrino mass $m_{\nu_1}$ have also been investigated.

The detailed plan of the paper is as follows. To set notations and
conventions as well as to make the paper self contained, in
Section (\ref{form}) we present some of the essentials regarding
texture 6 zero Dirac neutrino mass matrices. Inputs used in the
present analysis have been given in Section (\ref{in}). The
analysis pertaining to inverted, normal hierarchy and degenerate
scenario of neutrino masses have been respectively presented in
Sections (\ref{inv}), (\ref{nor}) and (\ref{deg}). Finally,
Section (\ref{summ}) summarizes our conclusions.

\section{Texture 6 zero Dirac neutrino mass matrices\label{form}}
In the Standard Model (SM), the mass terms corresponding to the
charged leptons and Dirac neutrinos having non zero masses are
respectively given by
\be
- \mathcal{L}_l= \overline{(l)}_L M_l (l)_R + h.c.  \ee and
\begin{equation}
-{\mathcal L}_D = \overline{(\nu_a)}_L M_{\nu D} (\nu_a)_R + h.c.,
\end{equation}
where L stands for left handedness, $M_l$ denotes the charged
lepton mass matrix, $M_{\nu D}$ is the complex $3\times 3$ Dirac
neutrino mass matrix and \be (\nu_a) \equiv \left( \ba {c} \nu_e
\\ \nu_{\mu} \\ \nu_{\tau} \ea
 \right),\qquad~~~~ (l) \equiv \left( \ba
{c} e \\ \mu \\ \tau \ea
 \right).\ee
The three flavor fields are $\nu_{aL}~ (a~=~e, \mu, \tau)$ and
$\nu_{aR}$ are the right-handed singlets which are sterile and do
not mix with the active neutrinos. The mass matrices $M_l$ and
$M_{\nu D}$ are arbitrary in the SM with a total of 36 real, free
parameters, these being quite large in number in comparison with
the 10 physical observables. Using the polar decomposition theorem
any general mass matrix M can be expressed as M=HU, where H
denotes a Hermitian and U a unitary matrix. In the present case,
the matrix U can be absorbed by redefining the right handed
singlet neutrino fields, therefore, enabling one to bring down the
number of free parameters from 36 to 18, which are further brought
down by considering textures, discussed below.

After defining the charged lepton and neutrino mass matrices,
their texture 6 zero Fritzsch structures are given as
 \be \label{frmm6}
 M_{l}=\left( \ba{ccc}
0 & A _{l} & 0      \\ A_{l}^{*} & 0 &  B_{l}     \\
 0 &     B_{l}^{*}  &  C_{l} \ea \right), \qquad
M_{\nu D}=\left( \ba{ccc} 0 &A _{\nu} & 0      \\ A_{\nu}^{*} & 0
&  B_{\nu}     \\
 0 &     B_{\nu}^{*}  &  C_{\nu} \ea \right),
 \ee
$M_{l}$ and $M_{\nu D}$ respectively corresponding to charged
lepton and Dirac neutrino mass matrices. It may be noted that each
of the above matrix is texture 3 zero type with $A_{l(\nu)}
=|A_{l(\nu)}|e^{i\alpha_{l(\nu)}}$ and $B_{l(\nu)} =
|B_{l(\nu)}|e^{i\beta_{l(\nu)}}$.

The formalism connecting the mass matrix to the neutrino mixing
matrix \cite{pmns} involves diagonalization of the mass matrices
$M_l$ and $M_{\nu D}$, details in this regard can be looked up in
\cite{ourreview}. In general, to facilitate diagonalization, the
mass matrix $M_k$, where $k=l, \nu D$, can be expressed as
\be
\label{mk}
 M_k= Q_k M_k^r P_k
 \ee
 or \be \label{mkr}
 M_k^r= Q_k^{\dagger} M_k P_k^{\dagger},
 \ee
 where $M_k^r$ is a real symmetric matrix with
real eigenvalues and $Q_k$ and $P_k$ are diagonal phase matrices
Diag$\{e^{i \alpha_{k}}, 1, e^{-i \beta_{k}} \}$ and Diag$\{e^{-i
\alpha_{k}}, 1, e^{i \beta_{k}} \}$ respectively. The real matrix
$M_k^r$ is diagonalized by the orthogonal transformation $O_k$,
e.g., \be M_k^{diag}= {O_k}^T M_k^r O_k \,, \label{mkdiag} \ee
 which on using equation (\ref{mkr}) can be
rewritten as
 \be M_k^{diag}= {O_k}^T Q_k^{\dagger} M_k
P_k^{\dagger} O_k \,. \label{mkdiag2} \ee
 The elements of the
general diagonalizing transformation $O_k$ can figure with
different phase possibilities, however these possibilities are
related to each other through the phase matrices \cite{ourreview}.
For the present work, we have chosen the possibility \be O_k=
\left( \ba{ccc} ~~O_k(11)& ~~O_k(12)& ~O_k(13)
\\
 ~~O_k(21)& -O_k(22)& ~O_k(23)\\
     -O_k(31) & ~~O_k(32) & ~O_k(33) \ea \right), \ee
where \beqn O_k(11) & = & {\sqrt \frac{m_{2} m_{3} (m_{3}-m_{2})}
     {(m_{1}-m_{2}+m_{3})
(m_{3}-m_{1})(m_{1}+m_{2})} } \nonum  \\ O_k(12) & = & {\sqrt
\frac{m_{1} m_{3}
 (m_{1}+m_{3})}
   {(m_{1}-m_{2}+m_{3})
 (m_{2}+m_{3})(m_{1}+m_{2})} }
\nonum   \\O_k(13) & = & {\sqrt \frac{m_{1} m_{2}
 (m_{2}-m_{1})}
    {(m_{1}-m_{2}+m_{3})
(m_{2}+m_{3})(m_{3}-m_{1})} } \nonum   \\ O_k(21) & = & {\sqrt
\frac{m_{1}
 (m_{3}-m_{2})}
  {(m_{3}-m_{1})(m_{1}+m_{2})} }
\nonum  \\O_k(22) & = & {\sqrt \frac{m_{2} (m_{1}+m_{3})}
  {(m_{2}+m_{3})(m_{1}+m_{2})} }
 \nonum    \\
O_k(23) & = & \sqrt{\frac{m_3(m_{2}-m_{1})}
 {(m_{2}+m_{3})(m_{3}-m_{1})} }
\nonum   \\O_k(31) & = &
 \sqrt{\frac{m_{1} (m_{2}-m_{1})
    (m_{1}+m_{3})}
{(m_{1}-m_{2}+m_{3})(m_{1}+m_{2})(m_{3}-m_{1})}} \nonum
\\O_k(32) & = & {\sqrt \frac{m_{2}(m_{2}-m_{1})
(m_{3}-m_{2})}{(m_{1}-m_{2}+m_{3})
 (m_{2}+m_{3})(m_{1}+m_{2})} }
 \nonum  \\
O_k(33) & = & {\sqrt \frac{m_{3}(m_{3}-m_{2})
(m_{1}+m_{3})}{(m_{1}-m_{2}+m_{3})
 (m_{3}-m_{1})(m_{2}+m_{3})}} \label{diageq} \,,
 \eeqn  $m_1$, $-m_2$,
$m_3$ being the eigenvalues of $M_k$. It may be added that without
loss of generality, we can always choose phase of one of the mass
eigenvalue relative to the other two. For details, we refer the
reader to \cite{ourreview, liuzhou}.

In the case of charged leptons, because of the hierarchy $m_e \ll
m_{\mu} \ll m_{\tau}$, the mass eigenstates can be approximated
respectively to the flavor eigenstates as has been considered by
several authors \cite{xingn5, fuku5}. Using the approximation,
$m_{l1} \simeq m_e$, $m_{l2} \simeq m_{\mu}$ and $m_{l3} \simeq
m_{\tau}$, the first element of the matrix $O_l$ can be obtained
from the corresponding element of equation (\ref{diageq}) by
replacing $m_1$, $-m_2$, $m_3$ with $m_e$, $-m_{\mu}$, $m_{\tau}$,
e.g.,
 \be  O_l(11) = {\sqrt
\frac{m_{\mu} m_{\tau} (m_{\tau}-m_{\mu})}
     {(m_{e}-m_{\mu}+m_{\tau})
(m_{\tau}-m_{e})(m_{e}+m_{\mu})} } ~. \ee

In the case of neutrinos, for normal hierarchy of neutrino masses
defined as $m_{\nu_1}<m_{\nu_2}\ll m_{\nu_3}$, as well as for the
corresponding degenerate case given by $m_{\nu_1} \lesssim
m_{\nu_2} \sim m_{\nu_3}$, equation (\ref{diageq}) can also be
used to obtain the elements of diagonalizing transformation for
Dirac neutrinos. The first element can be obtained from the
corresponding element of equation (\ref{diageq}) by replacing
$m_1$, $-m_2$, $m_3$ with $m_{\nu 1}$, $-m_{\nu 2}$, $m_{\nu 3}$
and is given by
 \be O_{\nu D}(11)  =  {\sqrt \frac{m_{\nu_2} m_{\nu 3} (m_{\nu 3}-m_{\nu 2})}
     {(m_{\nu 1}-m_{\nu 2}+m_{\nu 3})
(m_{\nu 3}-m_{\nu 1})(m_{\nu 1}+m_{\nu 2})} }, \label{nhelement}
\ee where $m_{\nu_1}$, $m_{\nu_2}$ and $m_{\nu_3}$ are neutrino
masses.

In the same manner, one can obtain the elements of diagonalizing
transformation for the inverted hierarchy case defined as
$m_{\nu_3} \ll m_{\nu_1} < m_{\nu_2}$ as well as for the
corresponding degenerate case given by $m_{\nu_3} \sim m_{\nu_1}
\lesssim m_{\nu_2}$. The corresponding first element, obtained by
replacing $m_1$, $-m_2$, $m_3$ with $m_{\nu 1}$, $-m_{\nu 2}$,
$-m_{\nu 3}$ in equation (\ref{diageq}), is given by
 \be O_{\nu D}(11)  =  {\sqrt \frac{m_{\nu_2} m_{\nu 3} (m_{\nu 3}+m_{\nu 2})}
     {(-m_{\nu 1}+m_{\nu 2}+m_{\nu 3})
(m_{\nu 3}+m_{\nu 1})(m_{\nu 1}+m_{\nu 2})} }. \label{ihelement}
\ee As already mentioned, one can choose the sign of one
eigenvalue relative to the other two, therefore, to facilitate
calculations for the inverted hierarchy case we have chosen
$m_{\nu 1}$ to be positive and both $m_{\nu 2}$ and $m_{\nu 3}$ to
be negative. The other elements of diagonalizing transformations
in the case of neutrinos as well as charged leptons can similarly
be found.

After the elements of diagonalizing transformations $O_l$ and
$O_{\nu D}$ are known, the Pontecorvo-Maki-Nakagawa-Sakata (PMNS)
matrix \cite{pmns} can be obtained through the relation
\be
 U = O_l^{\dagger} Q_l P_{\nu D} O_{\nu D} \,, \label{mixreal} \ee
where $Q_l P_{\nu D}$, without loss of generality, can be taken as
Diag$\{e^{-i\phi_1},\,1,\,e^{i\phi_2}\}$. The parameters $\phi_1$
and $\phi_2$ are related to the phases of mass matrices, i.e.,
$\phi_1= \alpha_{\nu D}- \alpha_{l}$, $\phi_2= \beta_{\nu D}-
\beta_{l}$ and can be treated as free parameters.

\section{Inputs used for the analysis\label{in}}
In the present analysis, we have made use of the results of the
latest global three neutrino oscillation analysis carried out by
Fogli {\it et al.} \cite{fogli}. At 1$ \sigma $ C.L. the allowed
ranges of the various input parameters are
\be
 \Delta {\it m}_{21}^{2} = (7.32-7.80)\times
 10^{-5}~\rm{eV}^{2},~~~~
 \Delta {\it m}_{23}^{2} = (2.33-2.49)\times 10^{-3}~ \rm{eV}^{2},
 \label{deltamass1}\ee
\be
s^2 _{12}  =  (0.29-0.33),~~~
 s^2_{23}  =  (0.37-0.41),~~~
s^2 _{13} = (0.021-0.026), \label{angles1} \ee where $ \Delta {\it
m}_{ij}^{2}$ 's correspond to the solar and atmospheric neutrino
mass square differences and $ s_{ij}$ 's correspond to the sine of
the mixing angle $\theta_{ij}$ where $i,j=1,2,3$. At 3$ \sigma $
C.L. the allowed ranges are given as
\be
 \Delta {\it m}_{21}^{2} = (6.99-8.18)\times
 10^{-5}~\rm{eV}^{2},~~~~
 \Delta {\it m}_{23}^{2} = (2.19-2.62)\times 10^{-3}~ \rm{eV}^{2},
 \label{deltamass2}\ee
\be
s^2 _{12}  =  (0.26-0.36),~~~
 s^2_{23}  =  (0.33-0.64),~~~
s^2 _{13} = (0.017-0.031). \label{angles2} \ee

For the purpose of the calculations, the masses and mixing angles
have been constrained by the data given in the above equations. In
the case of normal hierarchy, the explored range for the lightest
neutrino mass corresponding to $m_{\nu_1}$ is taken to be
$0.0001\,\rm{eV}-1.0\,\rm{eV}$, essentially governed by the mixing
angle $s_{12}$ related to the ratio $\frac{m_{\nu_1}}{m_{\nu_2}}$.
For the inverted hierarchy case also we have taken the same range
for the lightest neutrino mass corresponding to $m_{\nu_3}$. It
may be mentioned that our conclusions remain unaffected even if
the range is extended further. In the absence of any constraint on
the phases, $\phi_1$ and $\phi_2$ have been given full variation
from 0 to $2\pi$.

\section{Inverted hierarchy of neutrino masses\label{inv}}
\begin{figure}[hbt]
  \begin{minipage}{0.45\linewidth}   \centering
\includegraphics[width=2.in,angle=-90]{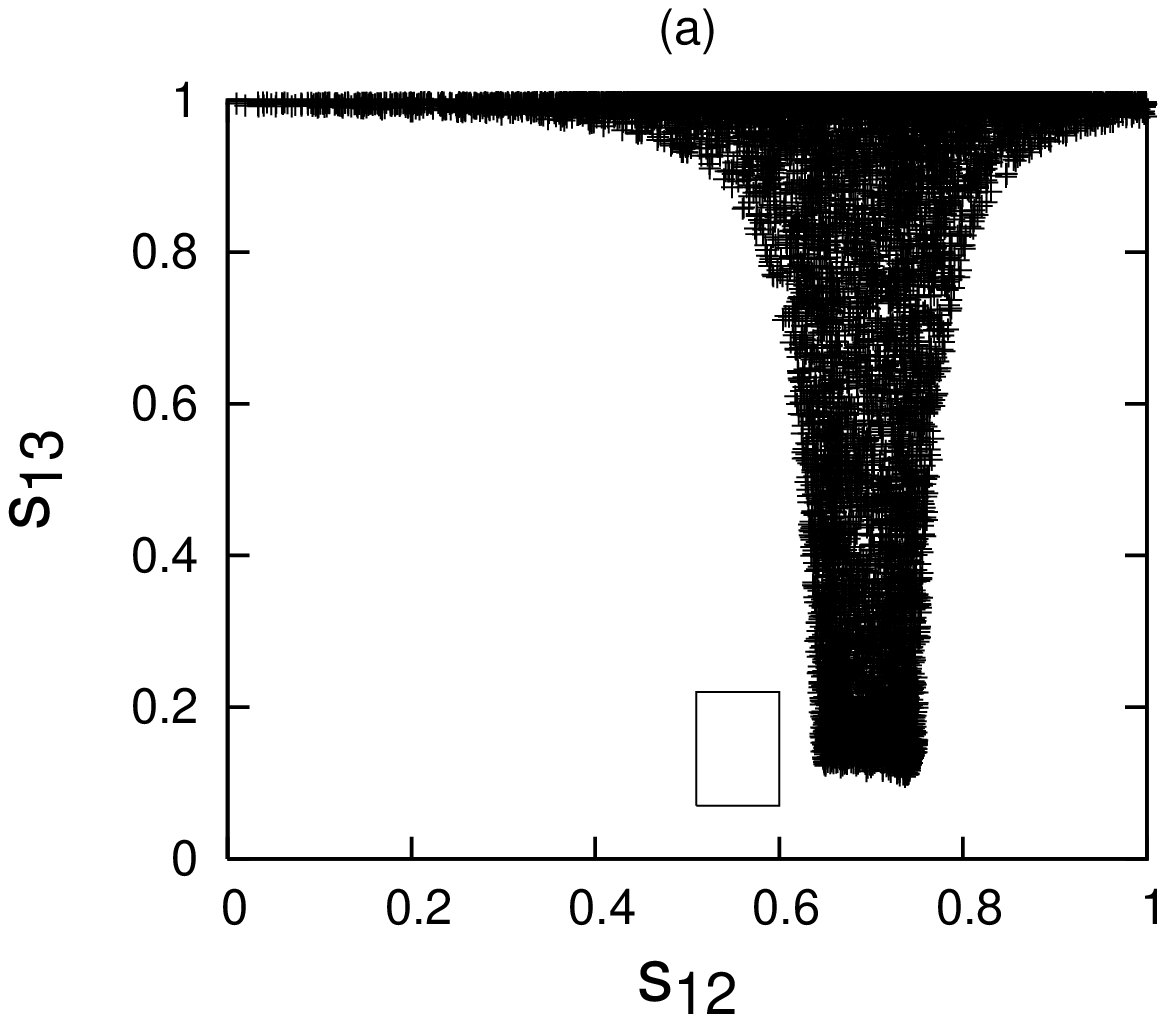}
    \end{minipage} \hspace{0.5cm}
\begin{minipage} {0.45\linewidth} \centering
\includegraphics[width=2.in,angle=-90]{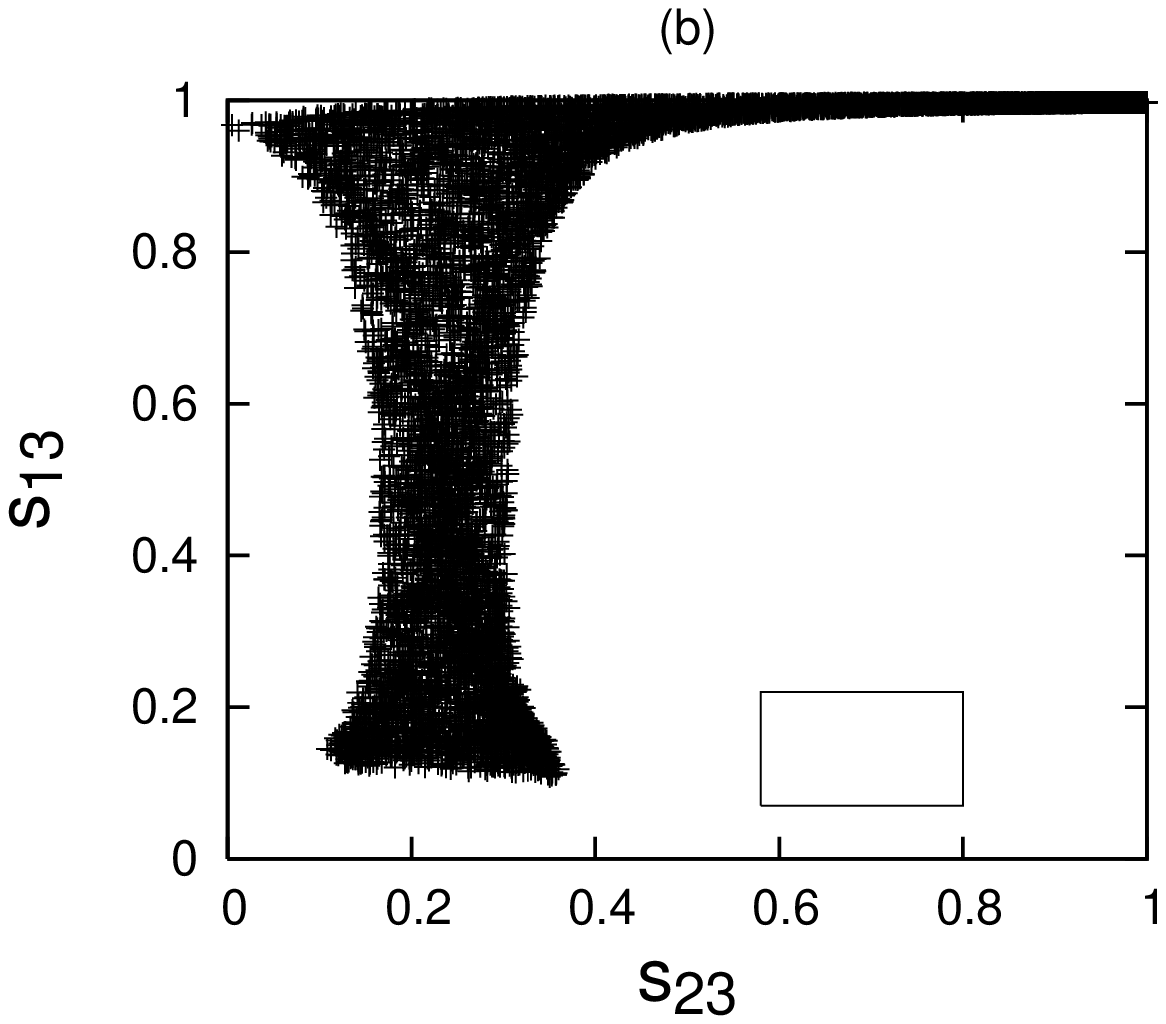}
  \end{minipage}\hspace{0.5cm}
   \caption{Plots showing the parameter space
corresponding to (a) $s_{13}$ and $s_{12}$ (b) $s_{13}$ and
$s_{23}$.}
  \label{fig1}
  \end{figure}

To examine the compatibility of texture 6 zero Dirac neutrino mass
matrices with the recent mixing data, we first discuss the
implications of mixing angle $\theta_{13}$ for the case pertaining
to inverted hierarchy of neutrino masses. To this end, in Figures
\ref{fig1}(a) and \ref{fig1}(b) we present the plots of the
parameter space corresponding to $s_{13}$ along with the other two
mixing angles $s_{12}$ and $s_{23}$ respectively. Giving full
allowed variation to other parameters, Figure \ref{fig1}(a) has
been obtained by constraining the angle $s_{23}$ by its
experimental bound given in equation (\ref{angles2}) and similarly
while plotting Figure \ref{fig1}(b) the angle $s_{12}$ has been
constrained by its experimental limits. Also included in the
figures are blank rectangular regions indicating the
experimentally allowed $3\sigma$ C.L. region of the plotted
angles. Interestingly, a general look at these figures reveals
that pertaining to inverted hierarchy of neutrino masses, the
texture 6 zero Dirac neutrino mass matrices are clearly ruled out
at 3$\sigma$ C.L.. This can be understood by noting that that the
plotted parameter space of the two angles has no overlap with
their experimentally allowed $3\sigma$ C.L. region.

\section{Normal hierarchy of neutrino masses\label{nor}}
\begin{figure}[hbt]
 \begin{minipage} {0.45\linewidth} \centering
\includegraphics[width=2.in,angle=-90]{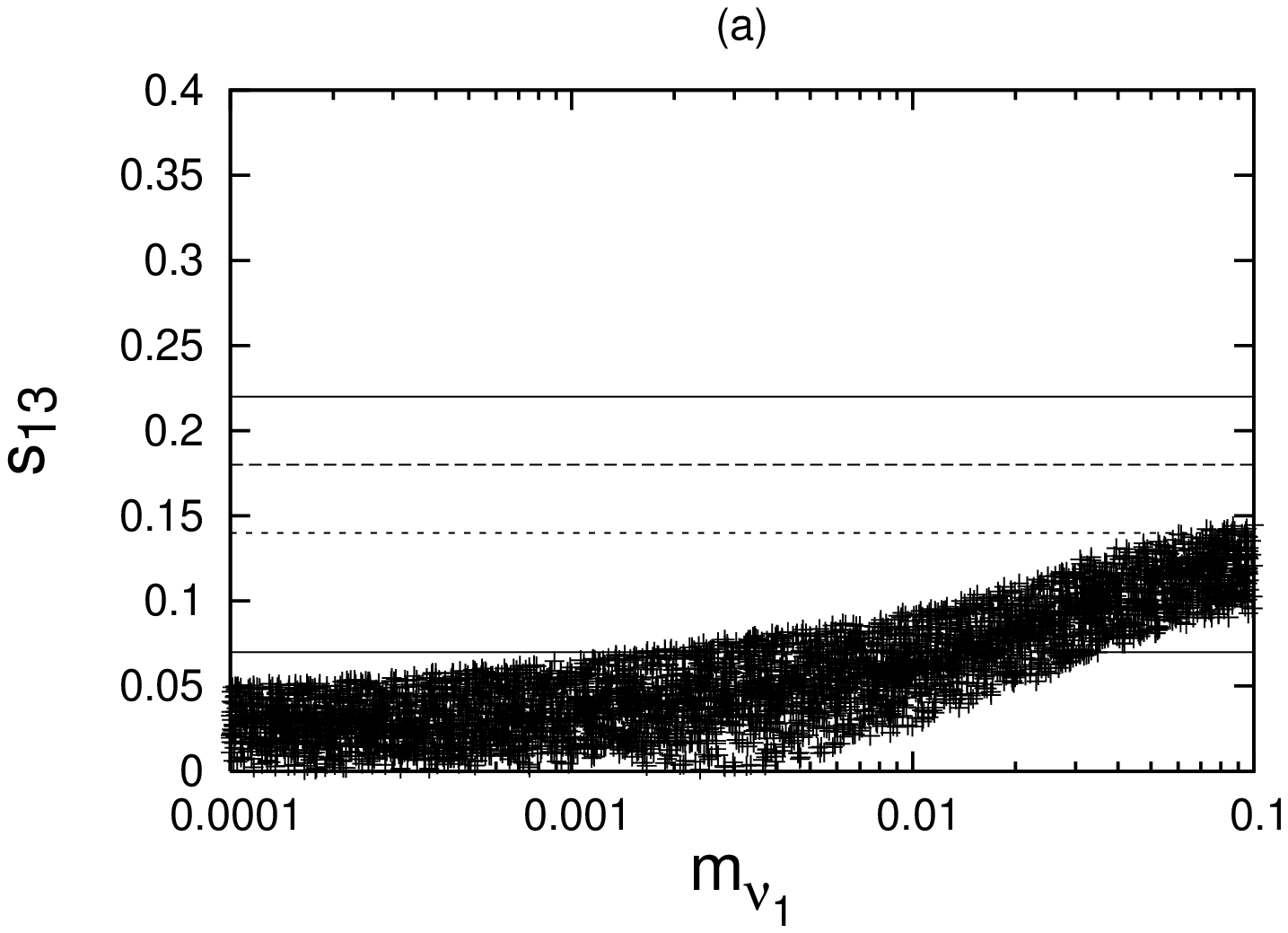}
  \end{minipage}\hspace{0.5cm}
\begin{minipage} {0.45\linewidth} \centering
\includegraphics[width=2.in,angle=-90]{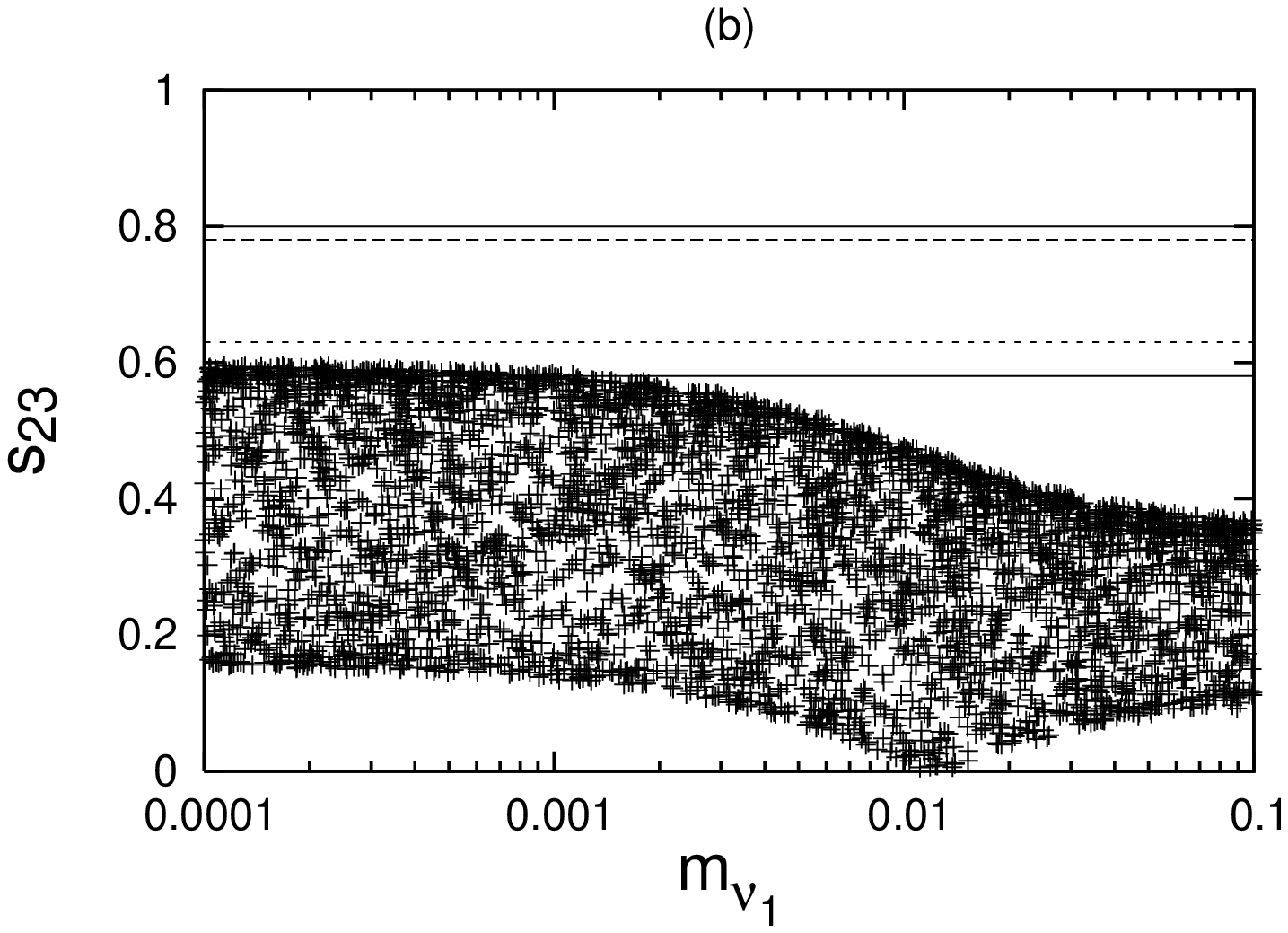}
  \end{minipage}
 \caption{Plots showing the variation of the lightest neutrino mass
$m_{\nu_1}$ with (a) $s_{13}$ and (b) $s_{23}$.} \label{fig2}
\end{figure}

After ruling out texture 6 zero Dirac neutrino mass matrices for
inverted hierarchy, we now examine the compatibility of these
matrices for the case of normal hierarchy. To this end, in Figure
\ref{fig2}(a) we present the graph of $s_{13}$ versus $m_{\nu_1}$,
in the graph the solid horizontal lines and the dashed lines
depict respectively the $3\sigma$ C.L. and $1\sigma$ C.L. range of
this angle. The graph depicts an interesting result that the
$1\sigma$ C.L. range of $s_{13}$ has no overlap with the plotted
values of the angle $s_{13}$ indicating towards the ruling out of
texture 6 zero mass matrices at $1\sigma$ C.L. for normal
hierarchy of neutrinos. However, a look at the figure also reveals
that corresponding to the $3\sigma$ C.L. range of $s_{13}$, one
gets get a lower bound on mass $m_{\nu_1}\sim 0.001 {\rm eV}$. One
may add that refinements in the measurement of angle $s_{13}$
would have interesting implications for this case.

To sharpen the above mentioned conclusions, in Figure
\ref{fig2}(b) we present the graph of angle $s_{23}$ w.r.t. mass
$m_{\nu_1}$, with the solid horizontal lines and the dashed lines
depicting respectively the $3\sigma$ C.L. and $1\sigma$ C.L. range
of this angle. Interestingly, from this figure one can conclude
that not only the $1\sigma$ C.L. range of $s_{23}$ again confirms
the ruling out of this case of texture 6 zero mass matrices, but
also corresponding to the $3\sigma$ C.L. range of this angle, one
finds that again the ruling out is largely confirmed. It may be
added in case we plot a graph of angle $s_{12}$ versus
$m_{\nu_1}$, it indicates towards compatibility of these mass
matrices with the data. However, it needs to be noted that to rule
out the matrices it is sufficient to do so from any one of the
mixing angle versus the mass $m_{\nu_1}$ graph.

\section{Degenerate scenario of neutrino masses\label{deg}}
The degenerate scenario of neutrino masses can be characterized by
either $m_{\nu_1} \lesssim m_{\nu_2} \sim m_{\nu_3} \sim
0.1~\rm{eV}$ or $m_{\nu_3} \sim m_{\nu_1} \lesssim m_{\nu_2} \sim
0.1~\rm{eV}$ corresponding to normal hierarchy and inverted
hierarchy respectively. As mentioned earlier, the diagonalizing
transformations for the above two cases are respectively the same
as the ones obtained for normal hierarchy of masses, equation
(\ref{nhelement}) and for inverted hierarchy of masses, equation
(\ref{ihelement}). Therefore, the conclusions regarding the
texture 6 zero Dirac neutrino mass matrices corresponding to both
normal and inverted hierarchy remain valid for this case also.

This can be understood from Figures (\ref{fig1}) and (\ref{fig2}).
While plotting Figures \ref{fig1}(a) and \ref{fig1}(b) the range
of the lightest neutrino mass is taken to be
$0.0001\,\rm{eV}-1.0\,\rm{eV}$, which includes the neutrino masses
corresponding to degenerate scenario, therefore by discussion
similar to the one given for ruling out texture specific mass
matrices for inverted hierarchy, these are ruled out for
degenerate scenario of neutrino masses as well. Similarly, for
degenerate scenario corresponding to normal hierarchy of neutrino
masses, Figure \ref{fig2}(b) clearly shows that the values of
$s_{23}$ corresponding to $m_{\nu_1} \lesssim 0.1~\rm{eV}$ lie
outside the experimentally allowed range, thereby ruling out the
mass matrices for degenerate scenario.

\section{Summary and conclusions\label{summ}}
To summarize, for Dirac neutrinos, we have carried out detailed
calculations pertaining to minimal texture characterized by
texture 6 zero Fritzsch-like mass matrices. Corresponding to
these, we have considered neutrino masses having normal, inverted
hierarchy as well as degenerate scenario. The compatibility of
these texture specific mass matrices have been examined by
plotting the parameter space corresponding to the recently
measured mixing angle $s_{13}$ along with the other two mixing
angles $s_{12}$ and $s_{23}$. Further, for the normal hierarchy
case, the implications of mixing angles on the lightest neutrino
mass $m_{\nu_1}$ have also been investigated.

Interestingly, the analysis reveals that using $1\sigma $
C.L.inputs, all the texture 6 zero cases of Dirac neutrino mass
matrices pertaining to normal, inverted hierarchy and degenerate
scenario of the neutrino masses seem to be completely ruled out,
for $3\sigma $ C.L. inputs, again these are largely ruled out.

\section*{Acknowledgements} P.F. would like to thank
University Grants Commission (Ref. No: F.
4-3/2006(BSR)/5-89/2007(BSR)) for financial support. G.A. would
like to acknowledge DST, Government of India (Grant No:
SR/FTP/PS-017/2012) for financial support. P.F., S.S., G.A.
acknowledge the Chairman, Department of Physics, P.U., for
providing facilities to work.

\end{document}